\documentclass[conf]{new-aiaa}
\usepackage[utf8]{inputenc}

\usepackage{graphicx}
\usepackage{amsmath}
\usepackage[version=4]{mhchem}
\usepackage{siunitx}
\usepackage{longtable,tabularx}
\usepackage{subcaption}
\title{Hacktivism Goes Orbital: Investigating NB65's Breach of ROSCOSMOS}

\author{Rajiv Thummala\footnote{PhD Student, Deptartment of Mech. and Aerospace Engineering, 124 Hoy Rd., Ithaca, NY, 14853} and Gregory Falco.\footnote{Assistant Professor, Department of Mech. and Aerospace Engineering, 124 Hoy Rd., Ithaca, NY, 14853}}
\affil{Sibley School of Mech. and Aerospace Engineering, Cornell University, Ithaca, 14850, USA}

\begin{document}

\maketitle

\begin{abstract}
In March of 2022, Network battalion 65 (NB65), a hacktivist affiliate of Anonymous, publicly asserted its successful breach of ROSCOSMOS's satellite imaging capabilities in response to Russia's invasion of Ukraine. NB65 disseminated a series of primary sources as substantiation, proclaiming the incapacitation of ROSCOSMOS's space-based vehicle monitoring system and doxing of related proprietary documentation. Despite the profound implications of hacktivist incursions into the space sector, the event has garnered limited attention due to the obscurity of technical attack vectors and ROCOSMOS's denial of NB65's allegations. Through analysis of NB65's released primary sources of evidence, this paper uncovers the probable vulnerabilities and exploits that enabled the alleged breach into ROSCOSMOS's ground and space segment. Additionally, we highlight lessons learned and the consequences this event has for the global aerospace community.

\end{abstract}

\section{Nomenclature}

{\renewcommand\arraystretch{1.0}
\noindent\begin{longtable*}{@{}l @{\quad=\quad} l@{}}
NB65  & Network Battalion 65 \\
CIDR & Classless Inter-Domain Routing \\
ASN & Autonomous System Number \\
OSINT & Open-Source Intelligence \\
RCE & Remote Code Execution \\
HTTP & Hypertext Transfer Protocol \\
SSL & Secure Sockets Layer \\
TLS & Transport Layer Security \\
XSS & Cross-Site Scripting \\
DDoS & Distributed Denial of Service \\
GNSS & Global Navigation Satellite System \\
LOG4J2 & Apache Log4j2 Logging Framework \\
DNS & Domain Name System \\
VMS & Vehicle Monitoring System \\
API & Application Programming Interface \\
CDHS & Command and Data Handling Subsystem\\
SWaP & Size, Weight, and Power \\
CISA & Cybersecurity and Infrastructure Security Agency \\
GIS & Geographical Information System \\

\end{longtable*}}

\section{Introduction}
Hacktivism, a portmanteau of "hack" and "activism", refers to the use of cyberattacks to advance a political or social agenda. Hacktivists--entities that engage in hacktivism--operate independently or in groups outside the scope of traditional government and military structures, often portraying themselves as populists. This form of activism has gained prominence in the digital age, with hacktivists breaching the security of adversary systems to draw attention and raise awareness to a cause. The anticipated escalation in the frequency of cyberattacks attributable to hacktivism is predicated on the diminishing technical barrier, largely driven through advents such as Cybercrime-as-a-Service (CaaS) and the democratization of penetration testing tools. This trend is further compounded by a growing reliance on digital infrastructure. Such a trajectory poses a significant threat to public and private sectors as the digital domain transforms into the primary stage for ideological expression and conflict.  

Various hacktivist groups have gained notoriety in recent times as a result of global conflicts. As recent as October 2023, reports emerged that hacktivists from the group "AnonGhost", who are seemingly conducting pro-Palestinian campaigns, have been launching DDoS attacks against Israel and are attempting to target infrastructure and APIs. The group went as far as to claim the alleged attack on the Israeli Red Alert missile warning platform \cite{newman_activist_nodate}. Hacktivism activity experienced a significant surge succeeding Russia's invasion of Ukraine in February of 2022 \cite{burgess_hacktivism_nodate}. Legacy hacktivist collective Anonymous was revitalized, but new groups were also formed. For instance. Ukraine's unprecedented IT Army, a volunteer group of hackers from around the world, has continuously launched DDoS attacks against Russian targets. Other hacktivist-linked groups have run doxing operations against Russian entities, leaking hundreds of gigabytes of proprietary data \cite{burgess_hacktivism_nodate}. 

In hacktivism, target selection is primarily driven by the potential for generating maximum public attention. In preceding decades, highly visited websites were a primary target. An early example is a website defacement attack against the U.S. Department of Justice (DOJ) web server that dates back to 1996. Protests were directed against the Communications Deceny Act (CDA), in regards to its provisions for screening offensive material online. Hacktivists reacted in a rebellious and provocative manner, adding obscene content to the homepage of the DOJ \cite{noauthor_virus_nodate}. As hacktivism evolved, however, critical infrastructure emerged as the optimal target. For instance, in 2012, the hacktivist group "Cutting Sword of Justice" carried out cyberattacks against critical systems of Saudi Aramco, the world's largest energy supplier and national oil company of Saudi Arabia, to protest Saudi Arabia's human rights record \cite{sadjadpour_irans_nodate}. Per a Waterfall Security report, in nearly every hacktivist incident in 2022, the sole motive was to disrupt critical infrastructure or services. Of the six total hacktivist attacks that occurred during the ongoing conflict between Iran and Israel or the Russo-Ukraine conflict, four incidents disrupted transportation operations and one targeted a steel mill which resulted in a fire and equipment damage. The last hacktivist attack targeted electric vehicle charging stations belonging to a power utility \cite{noauthor_high-impact_nodate}. 

The target shift to critical infrastructure is logical, given that disrupting these systems are likely to cause widespread panic and economic repercussions, drawing substantial attention to the hacktivist's cause. The symbolic significance of critical infrastructure enables hacktivists to underscore their contempt or political messages on a grand scale, enhancing their ability to convey their message and ultimately provoke change. Additionally, critical infrastructure is becoming increasingly interdependent, enabling hacktivists to cause cascading failures across multiple systems by exploiting a single attack vector. 

CISA identifies critical infrastructure as systems and assets, both physical and virtual, so vital to the United States that their incapacity or destruction would have a debilitating impact on national security, economic security, public health, or safety \cite{editor_critical_nodate}. This designation spans a broad spectrum, including sectors such as energy, transportation, water, healthcare, and communication. While the National Cybersecurity Standard and the Cyber Incident Reporting for Critical Infrastructure Act of 2022 mandate security practices and protocols for designated critical infrastructure by CISA, there exists a vulnerability in the realm of uncodified or undesignated critical infrastructure, which leave certain integral sectors or systems outside the prescribed security requirements. These systems, though not designated as critical infrastructure by CISA, could still possess the capability to cause catastrophic damage if breached. Space systems serve as a notable example of this phenomenon. While not officially designated as critical infrastructure by CISA, it is undeniable that various critical functionalities such as SATCOM, space-based nuclear radiation detection, and weather forecasting are dependent on space systems. Hacktivists may view space vehicles as potent symbols of technological reliance, making them attractive targets to amplify their messages or create widespread disruptions. The space domain, relatively uncharted territory for cyber-defense and unregulated by CISA, provides hacktivists with an environment where vulnerabilities may be more exploitable, and attacks may have a higher likelihood of success and attention. 

In this work, we analyze and document the first potential instance of hacktivism against satellites in history. On March 1st, 2022, hacktivist collective Network Battalion-65 claimed to have successfully executed a cyberattack against The State Corporation for Space Activities (ROSCOSMOS), the Russian equivalent of NASA, in response to the Russia's invasion of Ukraine. NB65 proceeded to release various screen-captures as evidence, but did not reveal attack vectors nor substantive technical details. Media outlets such as Newsweek reported that "it is unclear exactly how NB65 carried out the attacks" and that they could not verify them \cite{reporter_roscosmos_2022}. 

Through open-source analysis of NB65's released screen captures, we identified a vulnerability within ROSCOSMOS's ground segment and the corresponding exploit that likely facilitated NB65's execution of the alleged attack. Leveraging this information, we present an estimated cyber kill chain that unveils the sequential steps taken by NB65 to breach ROSCOSMOS's space-based vehicle monitoring system. Our findings address the previously limited technical details published by media outlets and security advisories following this event, contributing valuable insight into the rapidly evolving threat landscape of cyberattacks in the space domain.

\subsection{Network Battalion-65}
Network Battalion 65, also recognized as NB65, operates as a hacktivist group affiliated with the decentralized Anonymous collective. The organization has garnered recent attention for its activities directed at Russian state-owned and affiliated institutions, in response to the Russian invasion of Ukraine. Some of their operations include breaches into IP cameras and several SCADA systems. Beyond attacks on critical infrastructure, NB65 has become notorious for targeting Russian organizations with data leak operations and ransomware attacks \cite{labs_hacktivist_2022}. In late March of 2022, NB65 allegedly attacked the All-Russia State Television and Radio Broadcasting Company (VGTRK), resulting in the encryption of 786.2 GB of data, comprising 900,000 emails and 4,000 files \cite{horne_release:_2022}. Other Russian organizations NB65 has attacked include the document management operator Tensor, gas pipeline construction company SSK Gazregion LLC, and payment processor Qiwi \cite{noauthor_hackers_nodate}. 

\section{Primary Source Analysis}
On March 1st, 2022, NB65 asserted on X (formerly known as Twitter) that they had compromised ROSCOSMOS's satellite imaging capabilities, including a vehicle monitoring system. As featured in figure 1, NB65 claimed to have "deleted the WS02, rotated credentials, and shut down a server". 

"WSO2" in this context likely refers to the open-source integration platform that provide solutions for API management, identity and access management (IAM), and other middleware functionalities \cite{noauthor_what_nodate}. The utility of a WSO2 in the context of controlling a satellite would likely be for integrating various IT and software services in the ground segment. While further context is not provided (i.e.: redundancy measures, configuration, use-case, etc.), it can be conjectured that deleting the WSO2 would likely disrupt the satellite's ability to manage APIs, identities, or other integration services. The inability to manage APIs is likely to result in the disabling of IPC within the satellite. Embedded systems such as space vehicle's are especially dependent on IPC in contrast to conventional systems due to need for integrating heterogeneous components. Ensuring communication between these components is needed to remediate SWaP constraints and enable multi-agent coordination for a specific task. Breaching the WSO2 would therefore likely result in the compartmentalization of software and disable various critical functions for the satellite. Furthermore, configuration settings, security policies, and other elements related to API management would be lost. 

In the context of a ground station, IAM would involve controlling and managing user access to various resources within the satellite's network, CDHS, payloads, and/or computing environment. If IAM was breached as a result of deleting the WSO2, the satellite would likely lose its ability to verify the identities of users or other systems, leading to potential security vulnerabilities and cyberattacks. Additionally, access controls and permissions might be compromised, allowing unauthorized entities to gain access to sensitive information or functions. In the context of a VMS, this could range from gleaning sensitive payload information to performing unauthorized tasking. 

The rotation of credentials refers to a security practice where passwords, API keys, or other access credentials are changed regularly to reduce the risk of unauthorized access. In this context, rotating credentials infers that NB65 changed the access credentials associated with the VMS, exacerbating the ability for ROSCOSMOS to regain control. Consequences of "shutting down the server" vary depending on its configuration. In general, servers utilized in ground segments perform and facilitate the storage, processing, and dissemination of data gleaned by the satellite.  Considering NB65's claims that they "deleted various files across the server" this would likely refer to stored and processed VMS data which would have been prepared to disseminate to customers. 

This approach is reminiscent of the February 2022 cyberattack on Viasat, which only occurred one month prior to the purported NB65 attack on ROSCOSMOS. Black hat actors targeting Viasat initially gained access to a management server and then a network operations server. Succeeding reconnaissance efforts, they gained accessed to Viasat's FTP server (used for delivering software updates to modems) and deployed a wiper binary to purge flash memory across modems \cite{boschetti_space_2022}. The significance of this synchronicity is discussed in section VI.
\begin{figure}
    \centering
    \includegraphics[width=1\textwidth]{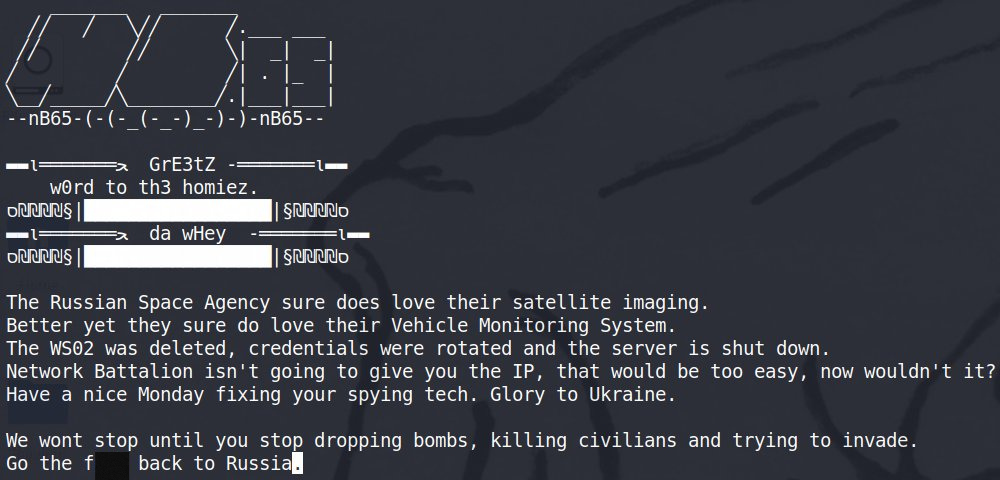}
    \caption{NB65's Manifesto (March 1st, 2022) \cite{noauthor_https://twitter.com/xxnb65/status/1498563301525102594_nodate}} 
    \label{fig: }
\end{figure}

\begin{figure}
  \begin{subfigure}{0.5\textwidth}
    \includegraphics[width=\linewidth]{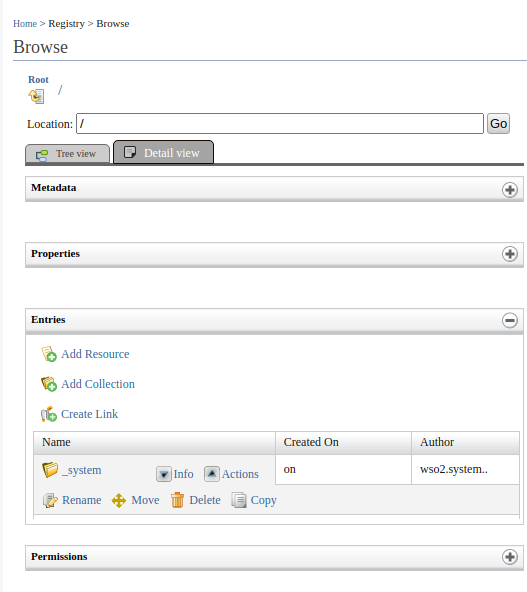}
    \caption*{Fig. 2 WSO2 Identity Server Registry Browser \cite{noauthor_x_nodate}}
    \label{fig:subfig1}
  \end{subfigure}%
  \begin{subfigure}{0.5\textwidth}
    \includegraphics[width=\linewidth]{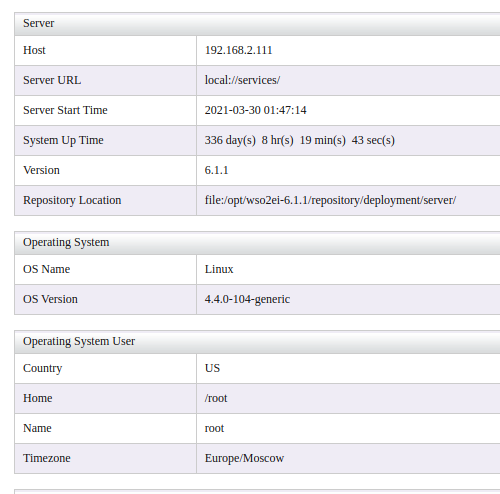}
    \caption*{Fig. 3 WSO2 Enterprise Integrator Home \cite{noauthor_x_nodate-1}}
    \label{fig:subfig2}
  \end{subfigure}

  \medskip

  \begin{subfigure}{\textwidth} 
    \centering
    \includegraphics[width=0.8\textwidth]{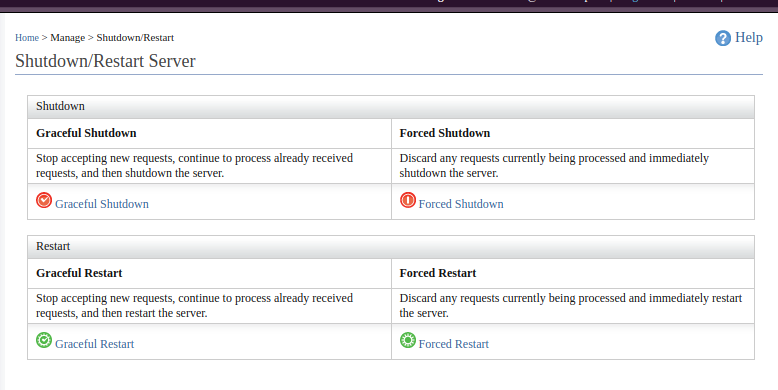}
    \caption*{Fig. 4 WSO2 Identity Server Management Console - Shutdown/Restart \cite{noauthor_x_nodate-2}}
    \label{fig:subfig3}
  \end{subfigure}

  \label{fig:overall}
\end{figure}

The screen-captures featured in figure's 2, 3, 4, 5, and 6 were released shortly after figure 1. Based on each of the user interfaces, it can be asserted that Roscosmos was employing WSO2's identity server management software and enterprise integrator. Figure 2, which appears to exhibit access to the interface of the WSO2 registry browser, could lead to severe consequences considering NB65 had write access. The registry browser is a critical component of the WSO2 Identity Server, as it provides the ability to view contents of the registry and perform various operations such as creating, editing, and deleting resources. Various attacks that NB65 could have executed include stealing user credentials, modifying access permissions and configuration settings, malware deployment, and installation of backdoors to enable persistence. In regards to the space segment, if the registry browser is configured to store predefined instructions for commands, NB65 could manipulate these instructions to perform malicious uplinks. 

Figure 3 initially appears to be either the home screen of the WSO2 identity server or enterprise integrator as they share the same user interface. Based on the repository location, however, it can be asserted that this is an enterprise integrator due to the presence of "wso2ei" in the '/opt' directory. The WSO2 enterprise integrator is an integration platform that helps enterprises connect disparate systems, applications, and data sources. Utility for this software in the ground segment would likely be for linking satellite telemetry systems (real-time data collection and routing), command and control systems, and enterprise resource planning systems. The enterprise integrator could also have been used to transform data into a format that can be easily interpreted and analyzed by other applications. The 6.1.1 version provides key insight into the vulnerability that likely enabled this attack. According to the WSO2 support matrix, WSO2 Enterprise Integrator 6.1.1 was released on May 04, 2017. The latest version of WSO2 Enterprise Integrator, 7.1.0, was released on August 12, 2020, more than a year and a half before the NB65 attack. Failure to update to the latest version of WSO2 would have exposed ROSCOSMOS to vulnerabilities, which is discussed at the end of this section. The operating system user section is assumed to be the main data NB65 sought to parade with this screen-capture. Both the home directory and user account appear to have root access privileges, indicating that NB65 had administrator-level control over the WSO2. Thus, this suggests that NB65 possessed the potential to execute a range of unauthorized operations within the functional scope of the WSO2 platform.

Figure 4 illustrates the server management console's shutdown/restart page, suggesting NB65's intention to convey their ability to initiate server shutdown or restart operations. The ramifications of shutting down or restarting the server include denial of service to the satellite, disruptions of processing capabilities, enabling malicious software updates, and disabling dependent 3rd-party assets and services. Figure 5 features a screen-capture of a supposed login screen for the Vehicle Monitoring System, which if authenticated could potentially enable tasking. \textit{Translations into English are denoted by double asterisks (**) and were provided by the authors of this paper, not NB65.} The implications of this capability would be menacing, ranging from the capability to induce kinetic effects (collisions, conjunctions, etc.), initiating re-entry, and installing malware (backdoors, rootkits, ransomware \cite{falco_wannafly:_2023}, spyware, etc.). It is also possible that this application is merely GIS software with no tasking capabilities, but could be exploited by altering image classification algorithms and gleaning tasking history as well as planned passes. The precise extent of damage potential is dependent on the capabilities of the VMS application.

\begin{figure}
    \begin{subfigure}{0.5\textwidth}
        \includegraphics[width=\linewidth]{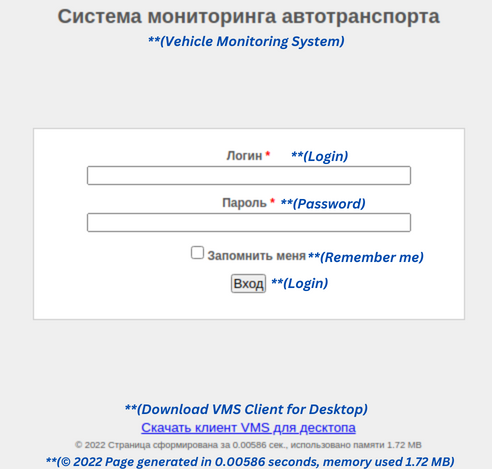}
        \caption*{Fig. 5 Vehicle Monitoring System \cite{noauthor_x_nodate-3}}
        \label{fig:vehicle-monitoring}
    \end{subfigure}%
    \begin{subfigure}{0.5\textwidth}
        \includegraphics[width=\linewidth]{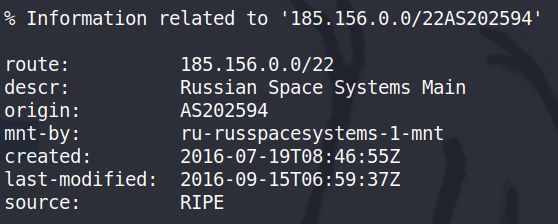}
        \caption*{Fig. 6 Internet Routing Registry \cite{noauthor_x_nodate-4}}
        \label{internet-routing}
    \end{subfigure}
\end{figure}

Figure 6 features metadata from an internet routing registry (IRR). The notation 185.156.0.0/22AS202594 denotes a specific IP address range using the CIDR format which appears to be associated with "Russian Space Systems Main". The IP address range begins at "185.156.0.0" and the "/22" indicates a subnet mask of 22 bits, defining the network portion of the address. This implies that the range spans from "185.156.0.0" to "185.156.3.255", encompassing various host addresses. Additionally, the notation includes the Autonomous System Number (ASN) "AS202594", signifying that these IP addresses are part of a collection of IP networks and routers operated by a single organization. The source is stated as "RIPE", which stands for Réseaux IP Européens and translates to "European IP Networks" in English. RIPE is one of the Regional Internet Registries (RIRs) worldwide \cite{feb_2023_internet_nodate}. These organizations are responsible for the allocation and registration of IP address space, the administration of ASNs, and other internet-related resources within their respective regions. This indicates that the information about the IP address has been sourced from RIPE NCC, and RIPE is the authoritative entity for managing that particular resource. Utilizing the IP address provided in the CIDR format, we performed an IP lookup to see if we could validate the authenticity of this information and glean more relevant material. Data from this lookup is depicted in figures 7 and 8. The listed domains were interpreted to be supplementary applications that ROSCOSMOS engages for various functionalities. We pinged these domains and they appeared to be reachable. 2 peers and upstream providers were found for the featured ASNs (not pictured) which were JSC Mastertel (AS29226) and LLC Nauka-Svyaz (AS8641), indicating that ROSCOSMOS was engaging these providers for internet connectivity or telecom. 2 specific uplinks were listed, which were Mastertel and RT-Comm (radiotelephone communication). 

Customers were found in the WHOIS report, which are other networks or entities that are connected to and receive services from the autonomous system. In this case, the sole customer was GTNT, a Russian telecommunications company that provides mobile satellite services in Russia. It is the official partner of Thuraya Telecommunications Company, a leading international mobile satellite operator. We discovered details regarding the domain owner, however, we deliberately excluded it from this paper for privacy concerns. NB65 is certain to have had further insight into the network and domain configurations of the ROSCOSMOS enterprise, indicating vulnerability to active and passive network reconnaissance.  

\begin{figure}
    \begin{subfigure}{0.5\textwidth}
        \includegraphics[width=\linewidth]{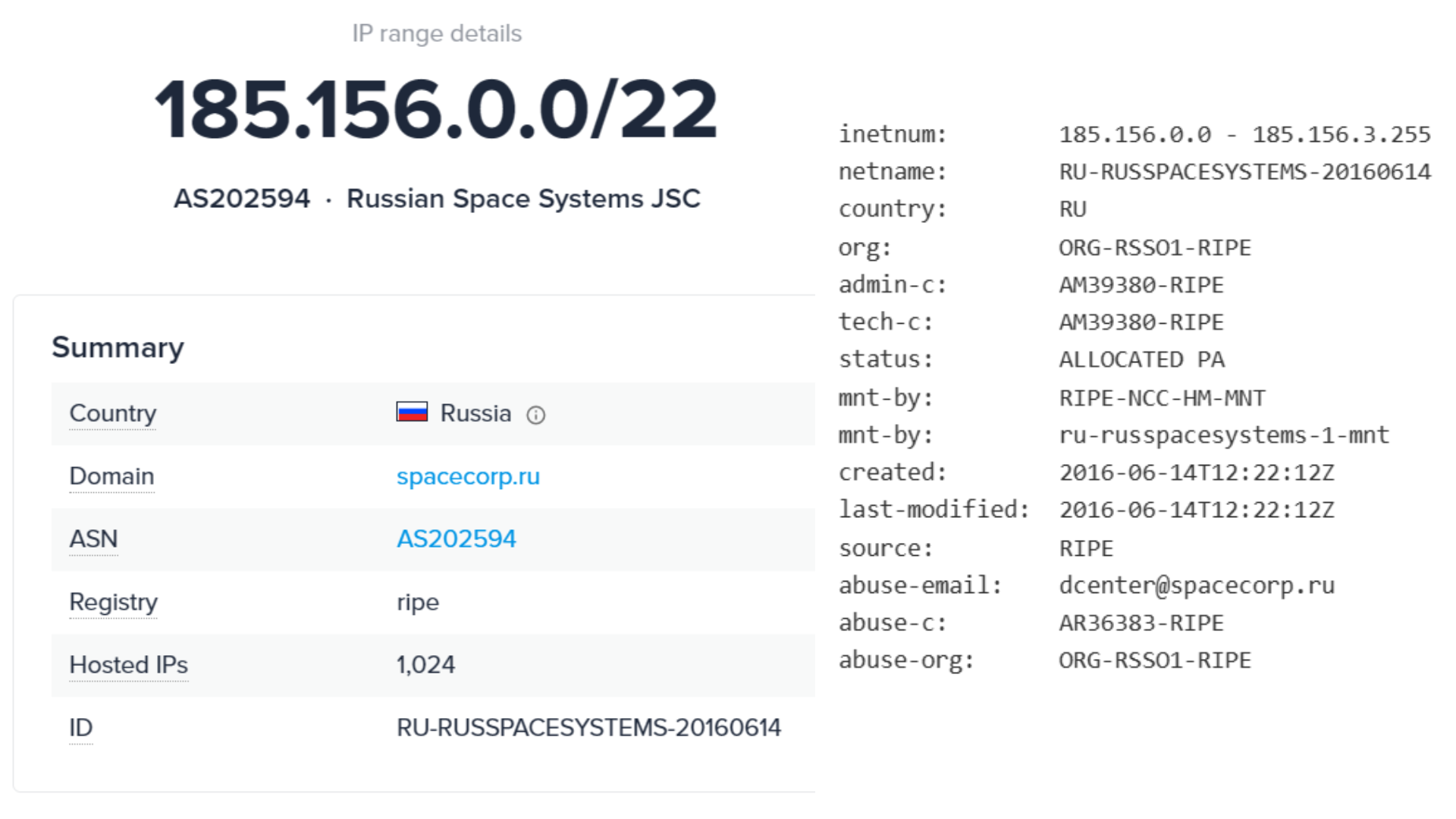}
        \caption*{Fig. 7 IP Lookup Part 1}
        \label{fig:vehicle-monitoring}
    \end{subfigure}%
    \begin{subfigure}{0.5\textwidth}
        \includegraphics[width=\linewidth]{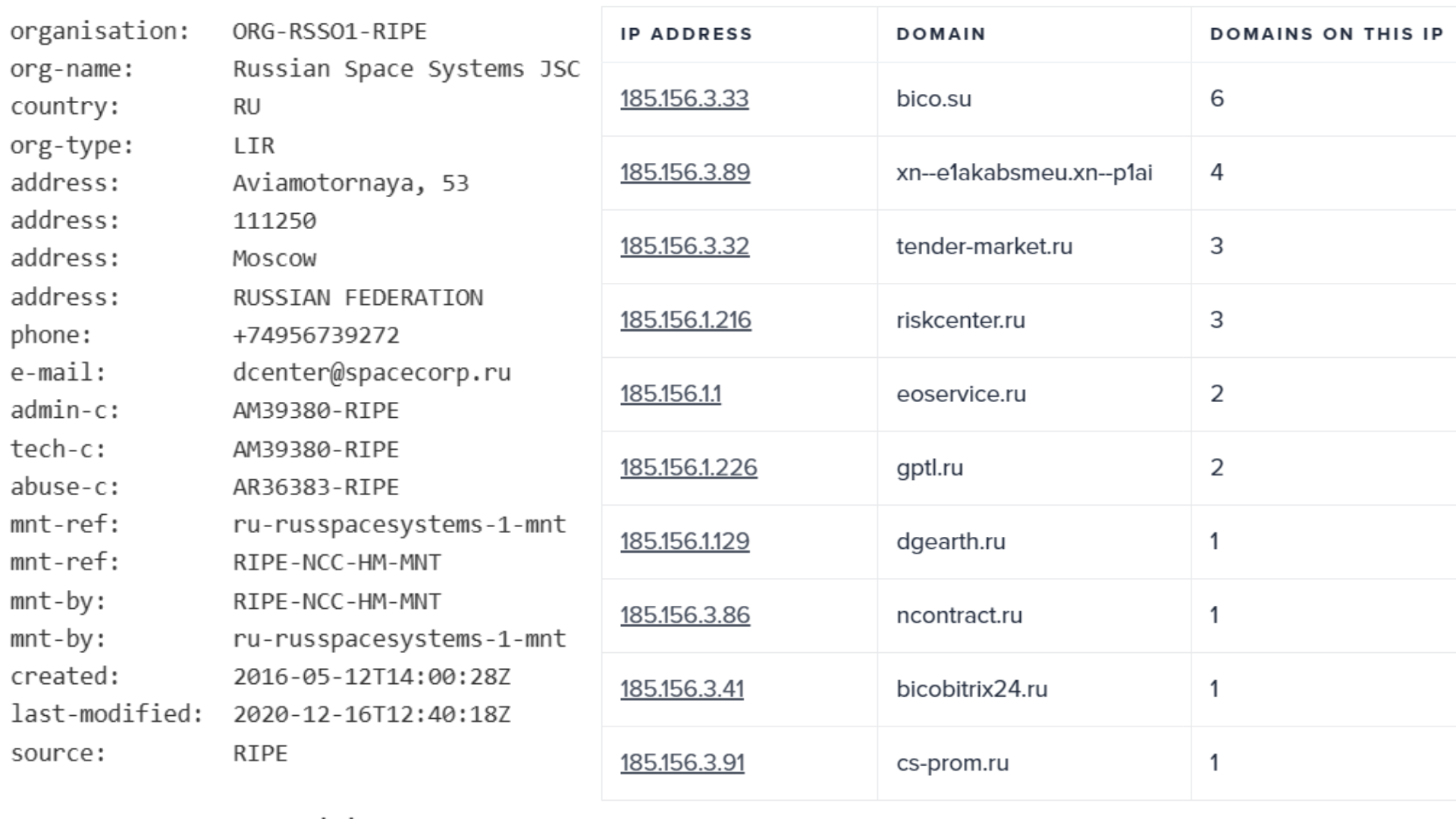}
        \caption*{Fig. 8 IP Lookup part 2}
        \label{internet-routing}
    \end{subfigure}
\end{figure}

On March 2nd, 2022, one day after NB65's claimed attack, Dmitry Rogozin--director general of Roscosmos (ret. July 2022)--took to X to refute NB65's claims. Rogozin claimed, "The information of these scammers and petty swindlers is not true. All our space activity control centers are operating normally \cite{noauthor_https://twitter.com/rogozin/status/1498903566135832577_nodate}." He followed up the post with an image of the Ukrainian actor Mikhail Vodyanoy in the 1967 film "Wedding in Malinovka" with the caption "Typical Ukrainian Hacker". Figures 9, 10, and 11 were published by NB65 on March 18th, 2022, 17 days after their initial publications, in response to Rogozin's claims on March 2nd. As seen in figure 9, NB65 retaliated to Rogozin's claims by doxing proprietary manuals, internal processes, and administration work. They also claimed to have leaked unreleased data from the Solar Observatory at ITSP SB RAS". The Solar Observatory ISTP SB RAS is a research facility of the Institute of Solar-Terrestrial Physics of Siberian Branch of Russian Academy of Sciences (ISTP SB RAS) located near the settlement of Mondy in Buryatia, Russia. It is one of the world’s leading observatories for studying solar activity and its effects on the Earth’s atmosphere and magnetosphere. The link under "targets" led to the main page of the ISTP SB RAS. 

Analyzing the target site, we found it to exhibit inadequate security measures, making it susceptible to various attacks. Despite prompting users to authenticate themselves, the site relied on HTTP without an SSL/TLS certificate, posing a potential risk to the confidentiality of user credentials. Additionally, the website's software was outdated, running on PHP version 7.4.6, and the PHP version was inadvertently disclosed through HTTP headers. Furthermore, the absence of an application firewall left the site vulnerable to DDoS and XSS attacks. It also lacked a security header to protect against clickjacking and content-type sniffing, and there was no content-security-policy directive in place. Our analysis of the site was passive and is therefore limited in scope. Based on the lack of rudimentary security protocols, the site is likely to be vulnerable to various OWASP Top 10 security risks. Given that these vulnerabilities were identified more than a year after NB65's publication of figure 9, we have high confidence that NB65 would have been able to breach this site. 

The loot mentioned in figure 9, supposedly of proprietary materials, could not be verified nor analyzed as the link is no longer valid. The notice states that either "the folder link has been removed as it violated our terms of service" or "the folder link has been disabled by the user". Whether NB65 took down the link or mega.nz identified genuine doxing and proceeded to take down the folder could not be determined. Figure 10 features what is interpreted to be the alleged loot. In addition to a folder of solar observatory data, there are installations and manuals for the VMS client which was also referenced in figure 5. This indicates that the solar observatory data may have been utilized by the vehicle monitoring system in some capacity. It is also possible that "VMS" in VMS client stands for another acronym such as Vendor Management System. 

Figure 11 features what is interpreted to be a breached Linux desktop. The file VMS-client.jar is a Java Archive (JAR) file and is likely a shortcut to the VMS client application pictured in figure 5. JAR files are used to store Java programs and libraries in a single file. Further substantiating this prediction is the \_start.bat file to the left of the VMS-Client.jar shortcut. The file "startup.bat" is typically a batch file utilized on Windows systems to execute a series of commands or operations during the startup of a program or system. In the context of Java applications, it is common to find "startup.bat" files associated with server applications, such as those based on Java EE (Enterprise Edition) or Java Servlet technology. This file may have been used to initiate the VMS-client server. Splash.jpg, based on standard naming convention, was employed for a splash page--also known colloquially as a loading screen. After extracting this logo from figure 11, we performed a reverse image search to identify its source. We were led to multiple sources, most of which when translated were titled “Russian Space Systems JSC”. The Department of State’s comments regarding this entity is highlighted as follows: “Joint Stock Company Rossiyskiye Kosmicheskiye Sistemy (Russian Space Systems JSC), a Russian space instrument building corporation, carries out activities to implement Russia’s state defense order. Russian Space Systems JSC is involved in Russia’s import substitution program in the context of Russia’s state defense order as well as associated space engineering activities. Russian Space Systems JSC has also been involved with Russian missile-related activities. As additional information, Russian Space Systems JSC has supported Russian government space systems that the Russian military uses to perpetrate its war against Ukraine” \cite{noauthor_targeting_nodate}. It can therefore be assumed that Russian Space Systems JSC partnered with ROSCOSMOS on the VMS. Main\_icon.png features the logo of the Research Institute of Precision Instruments (RI PI), a Russian company that manufactures antennas, transmitters, receivers and computing facilities. RI PI provides various services for spacecraft radio control and docking, data acquisition, satellite data processing, transmission of satellite data, radar monitoring of the Earth, and low-orbit satellite communications systems. We can assume that the VMS was also engaging RI PI's services for certain functionalities. Header.jpg features a logo with text that translates to "monitoring". The exact organization this image corresponds to could not be readily identified, however, the text on the left image translates to "KAYA" which may aid future investigations. It is possible that KAYA could be referencing KAYA Instruments \cite{noauthor_kaya_nodate}, which develops industrial vision and traffic inspection solutions. 

The most important information in this screen-capture is the log4j.properties file. The log4j.properties file is a configuration file for the Log4j2 logging framework and is utilized to define the behavior of Java-based applications. In this context, we assume the Java-based application to be the VMS application featured in figure 5. While the file is not inherently malicious, it can be used by attackers to exploit the Log4j2 RCE vulnerability and execute arbitrary code on vulnerable systems. The remaining information in the screen capture is mostly trivial and/or covered in other screen captures provided by NB65. This suggests that the log4j.properties file may have been the key element NB65 wanted to convey. The Log4j2 RCE vulnerability arose in December of 2021, only 3 months preceding NB65's claimed attack. In fact, according to WSO2's security and compliance documentation \cite{noauthor_log4j2_nodate}, a Log4j2 zero-day vulnerability notice (CVE-2021-44228 / CVE-2021-45046 / CVE-2021-45105) was published on December 13, 2021, stating that WSO2 had been impacted and that customer actions were required. Impacted WSO2 Products and Deployments included WSO2 Identity Server 5.9.0 and above and WSO2 Enterprise Integrator 6.1.0 and above. While the version of the Identity Server was not provided by NB65, the Enterprise Integrator version was listed in figure 3 as 6.1.1. This indicates that the Enterprise Integrator version ROSCOSMOS was utilizing was freshly vulnerable to the Log4j2 RCE vulnerability at the height of its exploitation. A temporary patch for the vulnerability was provided in the security and compliance documentation for affected versions in the December 13th notice. ROSCOSMOS likely failed to actively persist with patches for their WSO2 software, given that they were multiple versions behind (6.1.1 vs 7.1.0). Utilizing the Log4j2 RCE exploit, NB65 would have been able to gain root access, delete the WSO2, rotate credentials, and gain the ability to shut down the server. It is also possible that NB65 deployed their flagship variation of the Conti ransomware, which was utilized in various attacks succeeding the attack on ROSCOSMOS, to delete the WSO2 \cite{noauthor_hackers_nodate}. 

\begin{figure}
  \begin{subfigure}{0.5\textwidth}
    \includegraphics[width=\linewidth]{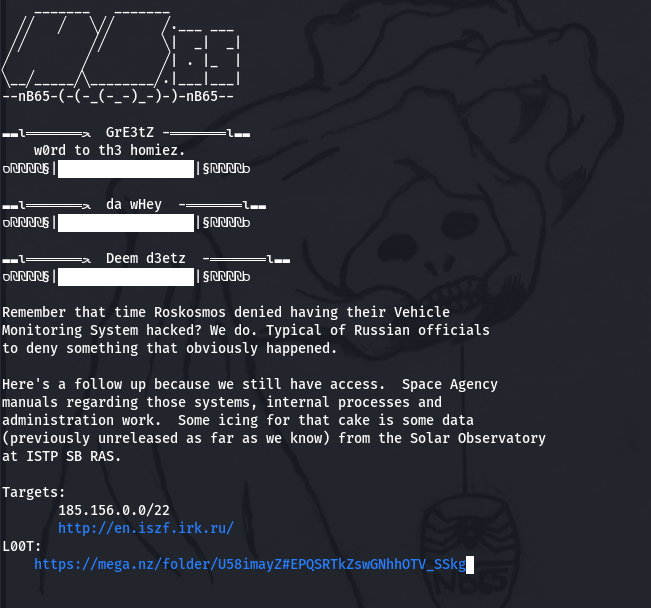}
    \caption*{Fig. 9 NB65 Response to Rogozin (March 18th, 2022) \cite{noauthor_https://twitter.com/xxnb65/status/1498563301525102594_nodate}}
    \label{fig:subfig1}
  \end{subfigure}%
  \begin{subfigure}{0.5\textwidth}
    \includegraphics[width=\linewidth]{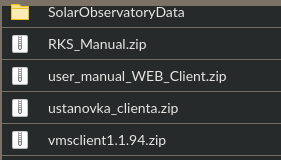}
    \caption*{Fig. 10 Purported Leaked ROSCOSMOS Materials \cite{noauthor_x_nodate-6}}
    \label{fig:subfig2}
  \end{subfigure}

  \medskip

  \begin{subfigure}{\textwidth} 
    \centering
    \includegraphics[width=0.8\textwidth]{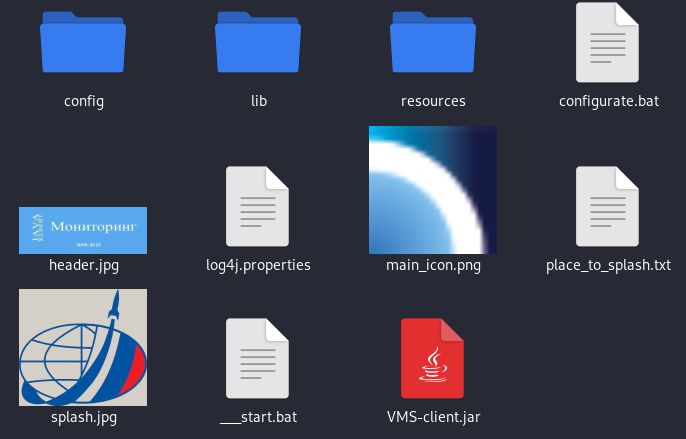}
    \caption*{Fig. 11 Breached ROSCOSMOS Desktop \cite{noauthor_x_nodate-7}}
    \label{fig:subfig3}
  \end{subfigure}
\end{figure}

\section{Cyber Kill Chain}
Based on the preceding analysis, an estimated cyber kill chain for NB65's breach of ROSCOSMOS is enumerated as follows: 

\subsection{Reconnaissance}
First, NB65 utilizes OSINT methods to glean information regarding ROSCOSMOS's network infrastructure, domain names, IP addresses, and subdomains. This information is obtained through DNS queries, WHOIS lookups, and other footprinting techniques to build a profile of ROSCOSMOS's online presence and ground segment. 

NB65 subsequently identifies open ports on a target system in the ground segment. Active and passive scans indicate that ROSCOSMOS is operating a VMS, is utilizing outdated WSO2 services, and is leveraging Apache Log4j2. This analysis is compounded with network mapping to architect ROSCOSMOS's IT environment, identify active devices, configurations, and providers.   

\subsection{Weaponization}
NB65 explores the potential deployment of the Log4j2 Remote Code Execution (RCE) exploit against the outdated WSO2 Enterprise Integrator. This effort is informed by the contemporary notoriety of the Log4j2 RCE exploit and NB65's awareness of the log4j.properties file used for logging within WSO2. Furthermore, security reports concerning WSO2's vulnerabilities and the imperative to patch the service had been circulating for 3.5 months prior to NB65's actions. Recognizing that the WSO2 Enterprise Integrator in use by ROSCOSMOS is susceptible to the Log4j2 RCE vulnerability, NB65 formulates a method to interact with the Log4j2 logging system and develops a payload to exploit the gained RCE capability. 

\subsection{Delivery}
NB65 engages the Log4j2 logging system, exploiting its vulnerability to the Log4j2 RCE exploit to gain unauthorized access to ROSCOSMOS's WSO2 environment. 

\subsection{Detonation}
NB65 performs unauthorized RCE to purge the WSO2, rotate credentials, shutdown the server, and escalate access for control over the VMS.  

\subsection{Installation}
Malware is installed to enable persistence, such as backdoors, rootkits, and spyware.

\subsection{Command and Control}
NB65 disrupts access to the space-based Vehicle Monitoring System (VMS), achieving this by compromising either the VMS client or application. Subsequently, they perform unauthorized data exfiltration, retrieving proprietary information such as manuals related to the VMS and solar observatory data.

\subsection{Act on Objectives}
NB65 publicizes the alleged success of their attack and provides various screen captures of the breach as evidence.

\section{Validity of Claims}
We acknowledge that there is a considerable lack of validity and reliability in NB65's claims and released evidence. It is possible that NB65 doctored the published screenshots to draw attention. This is especially true of items such as figure 4 which feature no personalized or proprietary information that can be related back to ROSCOSMOS. It is also important to reiterate that ROSCOSMOS denied the claims of NB65, stating that their systems were functioning as intended. According to Interfax news agency, Roscosmos head Dmitry Rogozin denied media reports that Russian satellite control centres had already been hacked amid Moscow's invasion of Ukraine, while warning against any attempts to do so. He further stated that "offlining the satellites of any country is actually a casus belli, a cause for war" \cite{noauthor_russia_2022}. However, we also note that Qiwi, an organization that NB65 had been proven to attack successfully, had also denied NB65's claims of attack \cite{arsdale_labs:_nodate}. Other points of contention are that NB65 claimed to have leaked Kasepersky's source code, but the so-called leak contained only information about Kaspersky that is or at some point was already publicly available \cite{lapienyte_long-awaited_nodate}. Despite the assurance of damage NB65 may or may not have inflicted, the consequences of this event for global space security are significant and are discussed in the following section.

\section{Consequences of this Event for Global Space Security}
NB65's attack on ROSCOSMOS is momentous, given that it is one of the first potential documented instances of hacktivism against satellites. Following this attack, various other hacktivist organizations have also sought to attack space-based assets such as the pro-Ukranian group OneFist, who claimed to have penetrated Gonets, a Russian low Earth orbit (LEO) satellite communications network, deleting a database that is crucial to its functioning \cite{petkauskas_we_nodate}. In April of 2023, a pro-Palestinian hacktivist group known as GhostSec claimed to have targeted Israel's satellite GNSS receivers \cite{ghost_israel_2023}. 

These incidents highlight a concerning trend as hacktivism expands its reach into the space sector, posing significant challenges to the  integrity and security of satellite systems. As the world becomes increasingly dependent on space-based technologies for critical functions, the space sector has become a prime target for malicious actors. The purported breach orchestrated by NB65 serves as a significant warning to aerospace organizations that they are not immune to the rising threat of hacktivism. While this incident raises concerns about fortifying ground-based infrastructure critical to space vehicles, it also introduces the potential for malicious actors to execute harmful uplinks with the capability to cause physical damage. While the validity of NB65's attack remains unconfirmed, performing the Log4j2 RCE exploit on a command and control system for a space vehicle could have catastrophic consequences unprecedented in the space sector.

This breach emphasizes the importance of securing all components of the space ecosystem from cyberattacks through active software updates, patch management, and comprehensive vulnerability assessments to identify and mitigate potential threats promptly. The compromise of third-party software, exemplified in this case with the WSO2 Enterprise Integrator and Identity Server, highlights the need for vigilance in managing dependencies in external software components and timely patching of vulnerabilities—a persisting issue in the space cybersecurity domain, as evidenced by the February 2022 attack on Viasat \cite{boschetti_space_2022}.

Redundancy measures must be implemented to ensure that critical functions offered by space vehicles can persist in the event of a cyberattack. This includes the integration of backup systems, alternative links, and air-gapped ground segments. As the world's reliance on space technologies continues to grow, it is essential for the space industry, governments, and cybersecurity experts to work collaboratively to fortify the security of space assets. This includes investing in research and development to create resilient and secure satellite systems \cite{finke_satellite_2023}, establishing international standards for space cybersecurity \cite{noauthor_ieee_nodate}, and promoting best practices for mitigating cyber threats in the space domain.

\bibliography{citationniner}

\begin{thebibliography}{33}
\newcommand{\enquote}[1]{``#1''}
\providecommand{\natexlab}[1]{#1}
\providecommand{\url}[1]{\texttt{#1}}
\providecommand{\urlprefix}{URL }
\expandafter\ifx\csname urlstyle\endcsname\relax
  \providecommand{\doi}[1]{\discretionary{}{}{}https://doi.org/#1}\else
  \providecommand{\doi}[1]{\discretionary{}{}{}\urlstyle{rm}\url{https://doi.org/#1}}\fi

\bibitem[{Newman(2023)}]{newman_activist_nodate}
Newman, L.~H., \enquote{Activist {Hackers} {Are} {Racing} {Into} the {Israel}-{Hamas} {War}—for {Both} {Sides},} \emph{Wired}, 2023.
\newblock \urlprefix\url{https://www.wired.com/story/israel-hamas-war-hacktivism/}.

\bibitem[{Burgess(2022)}]{burgess_hacktivism_nodate}
Burgess, M., \enquote{Hacktivism {Is} {Back} and {Messier} {Than} {Ever},} \emph{Wired}, 2022.
\newblock \urlprefix\url{https://www.wired.com/story/hacktivism-russia-ukraine-ddos/}.

\bibitem[{noa(2023{\natexlab{a}})}]{noauthor_virus_nodate}
\enquote{Virus {Bulletin} :: {VB2017} preview: {Hacktivism} and website defacement: motivations, capabilities and potential threats,} , 2023{\natexlab{a}}.
\newblock \urlprefix\url{https://www.virusbulletin.com/blog/2017/09/vb2017-preview-hacktivism-and-website-defacement-motivations-capabilities-and-potential-threats/}.

\bibitem[{Sadjadpour(2018)}]{sadjadpour_irans_nodate}
Sadjadpour, C.~A., Karim, \enquote{Iran’s {Cyber} {Threat}: {Espionage}, {Sabotage}, and {Revenge},} , 2018.
\newblock \urlprefix\url{https://carnegieendowment.org/2018/01/04/iran-s-cyber-threat-espionage-sabotage-and-revenge-pub-75134}.

\bibitem[{noa(2023{\natexlab{b}})}]{noauthor_high-impact_nodate}
\enquote{High-{Impact} {Attacks} {On} {Critical} {Infrastructure} {Climb} 140\%,} , 2023{\natexlab{b}}.
\newblock \urlprefix\url{https://securityintelligence.com/news/high-impact-attacks-on-critical-infrastructure-climb-140/}.

\bibitem[{Content(2023)}]{editor_critical_nodate}
Content, C., \enquote{critical infrastructure - {Glossary} {\textbar} {CSRC},} , 2023.
\newblock \urlprefix\url{https://csrc.nist.gov/glossary/term/critical_infrastructure}.

\bibitem[{Reporter(2022)}]{reporter_roscosmos_2022}
Reporter, E.~B., \enquote{Roscosmos {Head} {Rejects} {Anonymous} {Claim} {That} {Russian} {Satellites} {Were} {Hacked},} , Mar. 2022.
\newblock \urlprefix\url{https://www.newsweek.com/roscosmos-head-dmitry-rogozin-anonymous-russian-satellite-hack-1684033}.

\bibitem[{Labs(2022)}]{labs_hacktivist_2022}
Labs, F.~V., \enquote{Hacktivist {Attacks}: {Who} {They} {Target}, {What} {They} {Target} and {How},} , Dec. 2022.
\newblock \urlprefix\url{https://www.forescout.com/blog/the-increasing-threat-posed-by-hacktivist-attacks-an-analysis-of-targeted-organizations-devices-and-ttps/}.

\bibitem[{Horne(2022)}]{horne_release:_2022}
Horne, L.~B., \enquote{Release: {VGTRK} (786.2 {GB}),} , Apr. 2022.
\newblock \urlprefix\url{https://ddosecrets.substack.com/p/release-vgtrk-7862-gb}.

\bibitem[{noa(2022{\natexlab{a}})}]{noauthor_hackers_nodate}
\enquote{Hackers use {Conti}'s leaked ransomware to attack {Russian} companies,} , 2022{\natexlab{a}}.
\newblock \urlprefix\url{https://www.bleepingcomputer.com/news/security/hackers-use-contis-leaked-ransomware-to-attack-russian-companies/}.

\bibitem[{noa(2023{\natexlab{c}})}]{noauthor_what_nodate}
\enquote{What is {WSO2}?} , 2023{\natexlab{c}}.
\newblock \urlprefix\url{https://www.techtarget.com/searchapparchitecture/definition/WSO2}.

\bibitem[{Boschetti et~al.(2022)Boschetti, Gordon, and Falco}]{boschetti_space_2022}
Boschetti, N., Gordon, N.~G., and Falco, G., \enquote{Space {Cybersecurity} {Lessons} {Learned} from the {ViaSat} {Cyberattack},} \emph{{ASCEND} 2022}, American Institute of Aeronautics and Astronautics, Las Vegas, Nevada \& Online, 2022.
\newblock \doi{10.2514/6.2022-4380}, \urlprefix\url{https://arc.aiaa.org/doi/10.2514/6.2022-4380}.

\bibitem[{noa(2022{\natexlab{b}})}]{noauthor_https://twitter.com/xxnb65/status/1498563301525102594_nodate}
\enquote{https://twitter.com/{xxNB65}/status/1498563301525102594,} , 2022{\natexlab{b}}.

\bibitem[{noa(2022{\natexlab{c}})}]{noauthor_x_nodate}
\enquote{X,} , 2022{\natexlab{c}}.
\newblock \urlprefix\url{https://twitter.com/xxnb65/status/1498563590873300993/photo/1}.

\bibitem[{noa(2022{\natexlab{d}})}]{noauthor_x_nodate-1}
\enquote{X,} , 2022{\natexlab{d}}.
\newblock \urlprefix\url{https://twitter.com/xxnb65/status/1498563590873300993/photo/2}.

\bibitem[{noa(2022{\natexlab{e}})}]{noauthor_x_nodate-2}
\enquote{X,} , 2022{\natexlab{e}}.
\newblock \urlprefix\url{https://twitter.com/xxnb65/status/1498563590873300993/photo/3}.

\bibitem[{Falco et~al.(2023)Falco, Thummala, and Kubadia}]{falco_wannafly:_2023}
Falco, G., Thummala, R., and Kubadia, A., \enquote{{WannaFly}: {An} {Approach} to {Satellite} {Ransomware},} \emph{2023 {IEEE} 9th {International} {Conference} on {Space} {Mission} {Challenges} for {Information} {Technology} ({SMC}-{IT})}, 2023, pp. 84--93.
\newblock \doi{10.1109/SMC-IT56444.2023.00018}, \urlprefix\url{https://ieeexplore.ieee.org/document/10207354}, iSSN: 2836-4171.

\bibitem[{noa(2022{\natexlab{f}})}]{noauthor_x_nodate-3}
\enquote{X,} , 2022{\natexlab{f}}.
\newblock \urlprefix\url{https://twitter.com/xxnb65/status/1498563590873300993/photo/4}.

\bibitem[{noa(2022{\natexlab{g}})}]{noauthor_x_nodate-4}
\enquote{X,} , 2022{\natexlab{g}}.
\newblock \urlprefix\url{https://twitter.com/xxnb65/status/1498563748226809861/photo/1}.

\bibitem[{feb(2023)}]{feb_2023_internet_nodate}
\enquote{The {Internet} {Registry} {System},} , 2023.
\newblock \urlprefix\url{https://www.ripe.net/participate/internet-governance/internet-technical-community/the-rir-system/the-regional-internet-registry-system}.

\bibitem[{noa(2022{\natexlab{h}})}]{noauthor_https://twitter.com/rogozin/status/1498903566135832577_nodate}
\enquote{https://twitter.com/{Rogozin}/status/1498903566135832577,} , 2022{\natexlab{h}}.

\bibitem[{noa(2023{\natexlab{d}})}]{noauthor_targeting_nodate}
\enquote{Targeting {Russia}’s {Senior} {Officials}, {Defense} {Industrial} {Base}, and {Human} {Rights} {Abusers},} , 2023{\natexlab{d}}.
\newblock \urlprefix\url{https://www.state.gov/targeting-russias-senior-officials-defense-industrial-base-and-human-rights-abusers/}.

\bibitem[{noa(2023{\natexlab{e}})}]{noauthor_kaya_nodate}
\enquote{{KAYA} {Instruments} - {Everything} {You} {Need} for {Your} {Vision} {System},} , 2023{\natexlab{e}}.
\newblock \urlprefix\url{https://kayainstruments.com/}.

\bibitem[{noa(2021)}]{noauthor_log4j2_nodate}
\enquote{Log4j2 zero-day vulnerability ({CVE}-2021-44228 / {CVE}-2021-45046 / {CVE}-2021-45105),} , 2021.
\newblock \urlprefix\url{https://security.docs.wso2.com/en/latest/security-announcements/incident-clarifications/2021/log4j2-zero-day-vulnerability_CVE-2021-44228_CVE-2021-45046_CVE-2021-45105/#impact-on-wso2-products-and-deployments}.

\bibitem[{noa(2022{\natexlab{i}})}]{noauthor_x_nodate-6}
\enquote{X,} , 2022{\natexlab{i}}.
\newblock \urlprefix\url{https://twitter.com/xxNB65/status/1504937871882563584/photo/1}.

\bibitem[{noa(2022{\natexlab{j}})}]{noauthor_x_nodate-7}
\enquote{X,} , 2022{\natexlab{j}}.
\newblock \urlprefix\url{https://twitter.com/xxnb65/status/1504937871882563584/photo/2}.

\bibitem[{noa(2022{\natexlab{k}})}]{noauthor_russia_2022}
\enquote{Russia space agency head says satellite hacking would justify war -report,} \emph{Reuters}, 2022{\natexlab{k}}.
\newblock \urlprefix\url{https://www.reuters.com/world/russia-space-agency-head-says-satellite-hacking-would-justify-war-report-2022-03-02/}.

\bibitem[{Arsdale(2023)}]{arsdale_labs:_nodate}
Arsdale, C.~v., \enquote{From the {Labs}: {YARA} {Rule} for {Detecting} {NB65},} , 2023.
\newblock \urlprefix\url{https://www.reversinglabs.com/from-the-labs/from-the-labs-yara-rule-for-detecting-nb65}.

\bibitem[{Lapienytė(2023)}]{lapienyte_long-awaited_nodate}
Lapienytė, \enquote{Long-awaited {Kaspersky} leak doesn't seem to be a leak at all,} , 2023.
\newblock \urlprefix\url{https://cybernews.com/cyber-war/long-awaited-kaspersky-leak-doesnt-seem-to-be-a-leak-at-all/}.

\bibitem[{Petkauskas(2023)}]{petkauskas_we_nodate}
Petkauskas, \enquote{We breached {Russian} satellite network, say pro-{Ukraine} partisans,} , 2023.
\newblock \urlprefix\url{https://cybernews.com/cyber-war/we-breached-russian-satellite-network-say-pro-ukraine-partisans/}.

\bibitem[{{Ghost}(2023)}]{ghost_israel_2023}
{Ghost}, \enquote{Israel {Satellite} and water pumps {HACKED} {\textasciitilde}{GhostSec},} , Apr. 2023.
\newblock \urlprefix\url{https://telegra.ph/Israel-Satellite-and-water-pumps-HACKED-GhostSec-04-06}.

\bibitem[{Finke et~al.(2023)Finke, Thummala, Elbasheer, Hansen, Henry, Mamula, Noor, York, Zheng, and Falco}]{finke_satellite_2023}
Finke, J., Thummala, R., Elbasheer, R., Hansen, P., Henry, W., Mamula, D., Noor, A.~M., York, T., Zheng, K., and Falco, G., \enquote{Satellite {Cybersecurity} {Testbed} to {Improve} {Commercial} {Space} {Security},} \emph{{ASCEND} 2023}, American Institute of Aeronautics and Astronautics, Las Vegas, Nevada, 2023.
\newblock \doi{10.2514/6.2023-4768}, \urlprefix\url{https://arc.aiaa.org/doi/10.2514/6.2023-4768}.

\bibitem[{noa(2023{\natexlab{f}})}]{noauthor_ieee_nodate}
\enquote{{IEEE} {Standards} {Association},} , 2023{\natexlab{f}}.
\newblock \urlprefix\url{https://standards.ieee.org}.

\end{thebibliography}
\bibliographystyle{plain}
\end{document}